# Automated Brake Onset Detection in Naturalistic Driving Data

Shu-Yuan Liu[1*], Johan Engström[1], Gustav Markkula[2]

[1] Waymo LLC, 1600 Amphitheatre Parkway, Mountain View, California 94043 USA
[*]shuyuanl@waymo.com
[2] University of Leeds, Woodhouse Lane, Leeds, West Yorkshire LS2 9JT United Kingdom

## Abstract

Response timing measures play a crucial role in the assessment of automated driving systems (ADS) in collision avoidance scenarios, including but not limited to establishing human benchmarks and comparing ADS to human driver response performance. For example, measuring the response time (of a human driver or ADS) to a conflict requires the determination of a stimulus onset and a response onset. In existing studies, response onset relies on manual annotation or vehicle control signals such as accelerator and brake pedal movements. These methods are not applicable when analyzing large scale data where vehicle control signals are not available. This holds in particular for the rapidly expanding sets of ADS log data where the behavior of surrounding road users is observed via onboard sensors. To advance evaluation techniques for ADS and enable measuring response timing when vehicle control signals are not available, we developed a simple and efficient algorithm, based on a piecewise linear acceleration model, to automatically estimate brake onset that can be applied to any type of driving data that includes vehicle longitudinal time series data. We also proposed a manual annotation method to identify brake onset and used it as ground truth for validation. $R^2$ was used as a confidence metric to measure the accuracy of the algorithm, and its classification performance was analyzed using naturalistic collision avoidance data of both ADS and humans, where our method was validated against human manual annotation. Although our algorithm is subject to certain limitations, it is efficient, generalizable, applicable to any road user and scenario types, and is highly configurable.

## Introduction

Collision avoidance behavior modeling is a long-standing topic in applied traffic safety research (Markkula et al., 2016; Lee, 2008; Venkatraman et al., 2016; Markkula et al., 2012; Engström et al., 2024; Johnson et al., 2025; Schumann et al., 2025), especially in research on automated driving systems (ADS, SAE International, 2021). Collision avoidance behavior involves detecting the collision threat and executing the evasive maneuver with a sequence of actions to avoid a collision (Engström et al., 2024). Evasive maneuvers can be longitudinal (braking or accelerating), lateral (steering), or combinations of the two (Hankey et al., 2016). Response *timing* refers generally to the timing relations between events in a collision avoidance scenario whereas response *time* refers specifically to the time from a defined stimulus onset time until the onset of an evasive maneuver (Engström et al., 2024). Response timing and response time measurement has a wide range of use cases including but not limited to: 1) creating benchmarks of human drivers; 2) developing and evaluating ADS; 3) determining the impact of driver environmental factors on response timing (Engström et al., 2024). With the recent rapid advancement of ADS technology, response timing and response times become increasingly important in evaluation processes. For example, in recent papers, response times of a Non-Impaired road user with their Eyes ON the conflict (NIEON) was used to build human response benchmarks (Engström et al., 2024) and then used to evaluate ADS in collision avoidance scenarios (Scanlon et al., 2022).

Measuring response times is common in psychology, cognitive science, applied driving safety research, and other human behavioral research that involves human responses; however, the definition and determination of response onset vary. Particularly, in studies related to human driving behaviors in collision avoidance scenarios, driver response onset can, for example, be defined as an emotional ("oops") reaction to the event (e.g., a body movement or change in facial expression), a foot movement to reach for the brake pedal or a change in steering wheel angle (Hankey et al., 2016, Markkula et al., 2016, SAE International, 2015; Victor et al., 2015) or by changes in kinematic signals in longitudinal and/or lateral directions (Engström et al., 2024; SAE International, 2015). The term *reaction time* is sometimes used interchangeably with response time, but has also been more specifically defined as the time from "the onset of an initiating event to the *first observable response* to that event" (SAE International, 2015), while response time specifically refers to the reaction time plus a *movement time* defined as the time for "the responding foot or hand to move from one location to another." (SAE International, 2015; for brake onsets, the movement time is typically the time for the foot to move from the accelerator to the brake pedal). We here use the terms "response time" and "reaction time" based on the SAE J2944 definitions (SAE International, 2015). The annotation of physical reactions (e.g., "oops" reactions and/or body movement) is usually done by manually reviewing the in-vehicle camera video (Victor et al., 2015) which is time consuming and expensive. The onset of kinematic responses can be determined based on brake pedal and/or steering wheel signals. Driving simulator studies typically use these signals because they are readily available in the simulator output data (McGehee et al., 2000; Lee et al., 2021). However, in field studies, especially, naturalistic driving data, signals of pedal depression or steering wheel turning are only sometimes available for the ego vehicle (i.e., the instrumented vehicle) and not available at all for non-ego vehicles (i.e., vehicles that are interacting with the instrumented vehicles). There are studies that used the onset of brake lights as onset of braking responses (Green, 2000; SAE International, 2015); however, brake light is not always associated with evasive braking as the driver might be slowing down as an expected behavior (e.g., slowing down for an upcoming busy intersection or visible traffic queue ahead) before an actual braking response to a surprising and unexpected stimulus such as a sudden lane change by another vehicle in front of the ego (Engström et al., 2024; Victor et al., 2015). In naturalistic driving studies, response onsets are typically measured based on kinematic signals, in particular acceleration. Dingus et al. (2006) used deceleration onset to measure response onset but did not specify the exact measure. Recent naturalistic studies have used thresholds of kinematic signals to determine response onset (Jokhio et al., 2023; Terranova et al., 2024). Engström et al., (2024) used a manual method including visual inspection of acceleration signals and heuristic interpretation of the scenario to determine the evasive maneuver onset. In addition to the manually annotated first discernible physical reaction onset, Markkula et al. (2016) fit a piecewise linear model to the vehicle acceleration signal, to estimate the time when the subjective vehicle has begun to decelerate.

With the advancement of SAE Level 4 ADS (SAE International, 2021), driving logs of Level 4 ADS mileage that contain rich information of both the ADS, surrounding road users, and geographic information, are accumulated at a fast rate. For example, Waymo has accumulated over 56.7 million miles of rider-only self-driving miles through January 2025 (Kusano et al., 2025) and quickly reached 100 million miles in July 2025 (Waymo, 2025) , surpassing the mileage covered in the second Strategic Highway Research Program (SHRP2) (Dingus et al., 2015; Perez et al., 2016). The annotations in Naturalistic Driving Study (NDS) data such as eye glance and drivers' responses were done manually and, although accurate, can be time consuming and expensive (Klauer et al., 2011; Victor et al., 2015, Hankey et al., 2016). As Level 4 ADS companies continue to expand their operational design domain (ODD) and fleet size, the accumulation of mileage will be faster and the cost of manual evaluations will become more expensive. Thus, one of the challenges Level 4 ADS engineers and researchers are facing is how to analyze the data automatically, repeatably and efficiently. Modeling collision avoidance behaviors is of special interest in the development and evaluation of Level 4 ADS to establish human benchmarks and compare behaviors of humans and the automation (Scanlon et al., 2022; Engström et al., 2024; Schumann et al., 2025; Johnson et al., 2025). In Level 4 ADS log data, the vehicle control signals of other road users are not available. Therefore, the following gaps in the current definition and operationalization of response timing measures need to be addressed to enable efficient benchmarking and evaluation: 1) a consistent definition of response onset which is independent of vehicle control input signals; 2) an automated method for detecting response onsets on large scale dataset regardless of road user types, scenarios, and traveling speed; 3) a manual approach to determine response onset when the algorithm fails.

In this paper, we introduce an algorithm to estimate brake onset using only kinematic signals of other road users observed by Level 4 ADS perception system or from the Level 4 ADS driving trajectories. The algorithm can be applied to kinematics data from other sources such as driving simulator studies. For example, one use case would be if researchers want to compare response onsets between simulator data and NDS data where pedal depression information is not available.

While we recognize the importance of lateral evasive maneuvers (e.g., steering), this paper focuses on evasive braking maneuvers. This limitation will be further discussed in the discussion.

## Methods

### Definitions

Braking maneuvers can happen both in non-critical situations and as a collision avoidance behavior in traffic conflicts. The focus in the present paper is on the latter type of evasive braking maneuvers. We adopt the definition of a traffic conflict from ISO/TR 21974-1 (International Organization for Standardization, 2018):

*"Situation where the trajectory(ies) of one or more road users or objects (Conflict partners) led to one of three results: 1) a crash [Collision] or road departure, 2) a situation where an evasive maneuver was required to avoid a crash or road departure, or 3) an unsafe proximity between the conflict partners."*

Specifically, we focus on crashes and near-crashes among two conflict partners (referred to as agents in this paper) where collision avoidance behaviors were highly likely involved. The unsafe proximity type of conflicts are omitted intentionally because an evasive maneuver is not required to avoid a crash. Additionally, single vehicle conflicts, while potentially involving evasive maneuvers, are not addressed in this paper.

In traffic conflicts, evasive maneuvers include steering, braking, accelerating, or any combination of them (Hankey et al., 2016). Drivers have been reported to perform braking first and then steering when avoiding crashes if steering maneuver was involved (Li et al., 2019; Kaplan and Prato, 2012; Sarkar et al., 2021); however, this pattern is speed and scenario dependent where initial steering is more likely at higher speeds. Accelerating is a much less common type of collision avoidance behavior and not much studied in the literature. An important aspect of collision avoidance behaviors is that they are not premeditated, since they represent responses to unexpected and surprising conflicts (Hankey et al., 2016; Engström et al., 2024). For example, when approaching an intersection, the subject vehicle might initially slow down for the traffic ahead before braking hard because the lead vehicle unexpectedly slams on the brake for a yellow light. In this example, the hard brake after the initial slowing down is the collision avoidance behavior of the subject vehicle and the brake onset would be the beginning of the hard brake that is indicated by a sudden and sharp change in longitudinal acceleration.

In trajectory conflicts, the *initiator* is the road user who made the initial and unexpected maneuver(s) leading to the conflict. The other involved road user is referred to as the *responder*, who may or may not respond to the conflict with collision avoidance behaviors that might or might not successfully avoid a collision (Scanlon et al., 2021). It is likely that the responder did not exhibit any evasive maneuvers in near-crashes because the conflict was mitigated by the initiator for example by changing in its driving course. We did not exclude such events as there was still a projected trajectory overlap at some point during the interaction.

Prior research at determining the brake onset when distinct physical reaction and control signals are present laid the foundation for the current paper, particularly, the manual annotation of evasive maneuvers by Engström et al. (2024) and the piecewise linear model developed by Markkula et. al. (2016) that estimates deceleration onset. In the following sections, we discuss how the current study builds on and expands the prior studies.

## Datasets

A total of 219 conflicts that involved a Waymo rider-only (RO) car (i.e., Level 4 ADS operated without a human behind the wheel) and at least one other road user were analyzed. These events were randomly selected from all the trajectory conflicts observed in Waymo logs that are either a crash or a near-crash where a collision avoidance behavior was required to avoid a crash. Among the selected conflicts, the responder is either a Waymo vehicle or a non-Waymo other road user. For other road user responders, we included all observed agent types including: passenger vehicle drivers, motorcyclists, cyclists, micromobility device users, and walking pedestrians (see Table 1). Different conflict types such as front-to-rear, same direction lateral incursion and opposite direction lateral incursion were sampled (see Figure 1 and Table 2). The events were selected based on availability and they are not intended to be representative of the real-world distributions of agent or scenario types. All selected events took place on surface streets in dense urban areas, with various travelling speeds. The goal of this sampling approach is to demonstrate the universality and generalizability of the proposed method. The number of events by agent type and scenario type can be found in Table 1 and Table 2, respectively. Note that there is only one responder in each event, so the number of events equals the number of responders. The Waymo vehicles are Jaguar I-PACE, categorized as passenger cars.

Table 1. Number of crashes and near-crashes by agent type.

|  | Passenger cars | Motorcycles | Bicycles | Micromobility devices | Pedestrians (on foot) | All Agent Types |
| --- | --- | --- | --- | --- | --- | --- |
| **Total** | 201 | 10 | 5 | 1 | 2 | 219 |
| **Crashes** | 7 | 0 | 0 | 0 | 0 | 7 |
| **Near-crashes** | 194 | 10 | 5 | 1 | 2 | 212 |

Table 2. Number of crashes and near-crashes by scenario type. *Note 1, V2V stands for Vehicle to Vehicle and F2R stands for front-to-rear. Note 2, pedestrian scenario covers pedestrians who were either on foot or using micromobility devices.*

|  | V2V F2R | V2V Intersection | V2V Lateral | V2V Opposite Direction | Secondary Crash | Motorcycle | Cyclist | Pedestrian | All Scenarios |
|---|---|---|---|---|---|---|---|---|---|
| **Total** | 82 | 15 | 78 | 7 | 1 | 12 | 9 | 15 | 219 |
| **Crashes** | 0 | 2 | 2 | 1 | 1 | 0 | 1 | 0 | 7 |
| **Near-crashes** | 82 | 13 | 76 | 6 | 0 | 12 | 8 | 15 | 212 |

The conflict typology used to annotate scenario types is shown in Figure 1 (Kusano et al., 2023).

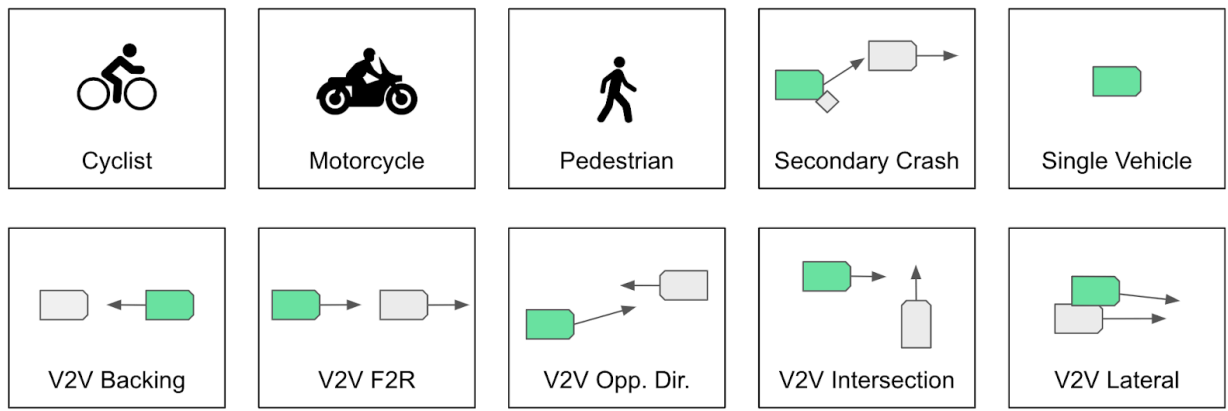

Figure 1. Conflicts types in the sample trajectory conflicts (green denotes ego vehicle).

## Manual ground truth annotation

Engström et. al (2024) annotated the brake onset of the responders based on acceleration time series plots. The annotation process used in this study is similar to Engström et. al (2024): we manually annotated the time point of brake onset using time series plots of longitudinal acceleration. There are two types of brake onset:

- If the agent was initially traveling at a constant speed or speeding up (i.e., no initial braking) prior to the conflict, the onset of braking maneuver is the brake onset.

- If the agent was initially slowing down (i.e., has initial braking) prior to the conflict, the sharp change in initial braking leading up to a harsher brake is the brake onset (see Figure 2).

In addition to the time series plot of kinematics, event video was used to provide contextual information. The contextual information is especially important when there is initial braking that is not in response to the unexpected conflict. In Figure 2, the agent was slowing down before and shortly after the conflict start time near t = 2.2 s before it started to hard brake near t = 3.8 s. The video confirms that the initial braking is premeditated as the responder was making a lane change and the much harsher braking starting around t = 3.8 s was in response to the unexpected braking of the lead vehicle.

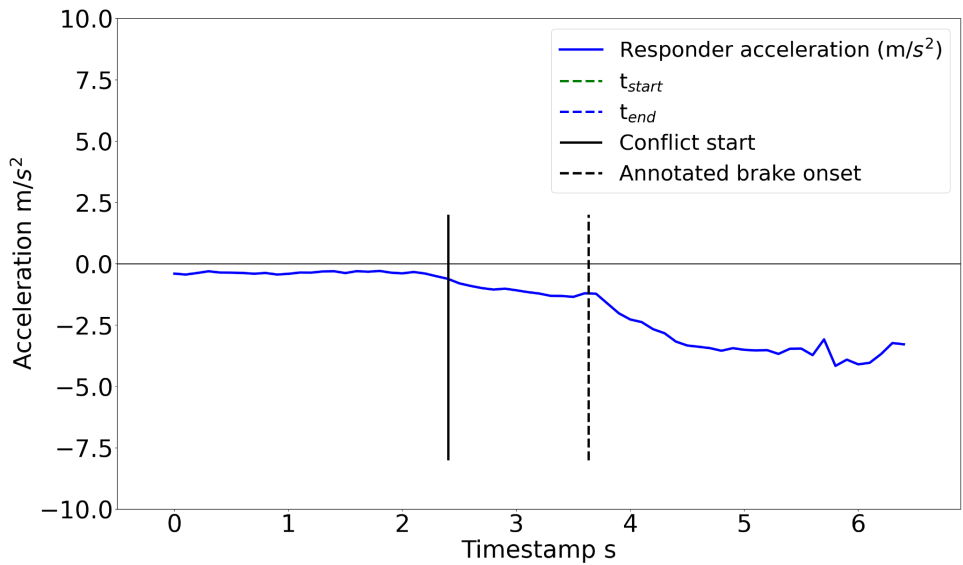

Figure 2. Example of a responder's braking response after initial slowing down that was not associated with the conflict.

The manual annotation of brake onset was used as the ground truth to evaluate the automatically estimated onset. A "no response" label was added if there was no braking maneuver or no response at all. A responder might not perform evasive maneuvers if the initiator changed its driving course and resolved the conflict. Therefore, the event could either be a crash or near-crash even if the responder did not respond.

The ground truth dataset also contains stimulus onset $T_1$ which is the first observable surprising evidence of the conflict and the start of the conflict and was manually annotated following a set of heuristic rules described in Engström et al., (2024). The $T_1$ annotation was here used to determine the time window during which the evasive maneuver might have occurred, as further described below.

Brake onset estimation with piecewise linear model

We developed a two-piece piecewise linear model (PLM) to fit on responders' longitudinal acceleration signals in traffic conflicts. The model contains three parameters:

$a_0$: the initial constant acceleration.

$t_B$: the time at which the acceleration starts to decrease.

$j_B$: the jerk at which the acceleration is decreasing.

The three parameters create two linear components that, pieced together, represent a braking reaction consisting of a single brake ramp (see Figure 3 top illustration).

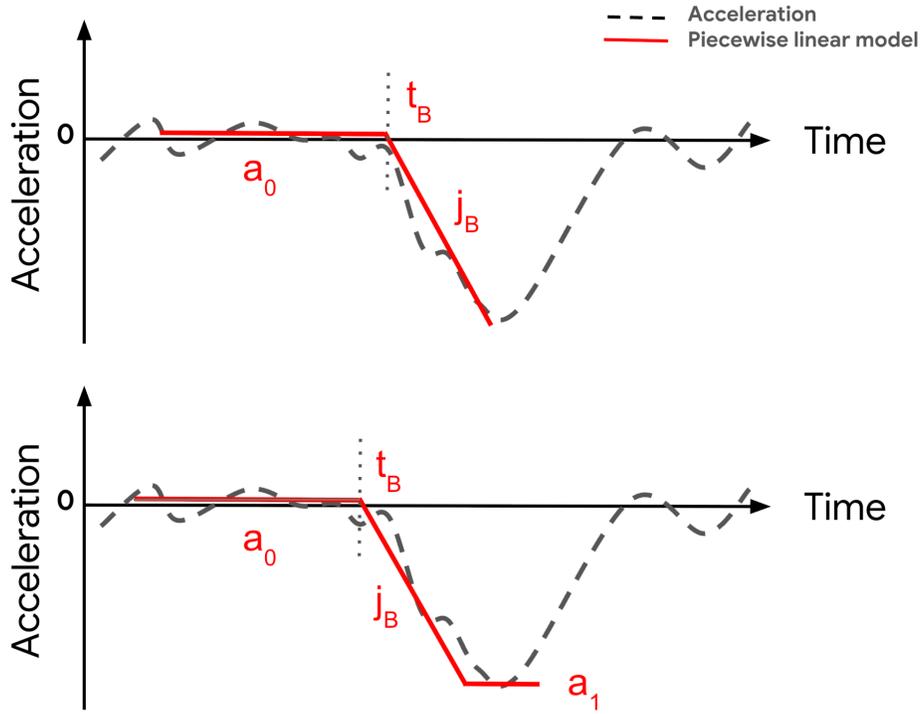

Figure 3. The illustration of the piecewise linear model fitting on the acceleration signal. The top figure is the two-piece model developed in the current paper. The bottom illustration is the original three-piece model in Markkula et. al (2016).

The model was based on the three-piece piecewise linear model used in Markkula et al. (2016) to analyze braking responses in rear-end crashes and near-crashes (See Figure 3 bottom plot). There are some key differences between the two models. The Markkula et al. (2016) model contains a fourth parameter $a_1$, or the final constant acceleration and fits on a time window that covers the entire braking maneuver as it aims to model the full braking profile. While we aim to optimize the estimation of $t_B$, or the brake onset with enhanced efficiency by eliminating $a_1$, using a customized grid search, and fitting on a time window that does not contain the entire braking maneuver. The details of the time window selection and grid search set up of the three-piece PLM can be found in Markkula et al. (2016) and that of our model is explained below.

The original PLM in Markkula et al. (2016) applied a limit on initial traveling speed as 15 km/h and used the same grid search set up of parameters for all events. The dataset used in this paper includes events with lower initial traveling speed. Thus, we customized the grid search for each event based on the specific kinematics. Our PLM (referred to as the model or algorithm in the rest of the paper) fits on the acceleration time series within a window of $[t_{start}, t_{end}]$. We define $t_{start}$ as $T_1 - 1$ s and $t_{end}$ as $T_{a\_min}$ which is the timestamp of minimum acceleration within $[T_1 - 1$ s, $T_1 + 4$ s$]$ for near-crashes or $[T_1 - 1$ s, 0.2 s before the crash$]$ for crashes. The 1 s and 0.2 s were selected based on observing the data. The model fit window was cut off shortly before the crash to eliminate the impact of collision on kinematics which can introduce large jerks and deceleration that might be confused as braking maneuvers and complicate the model fitting. We then define the search grid for each individual event as: $a_0 \in \{a_{max} - 1$ m/s$^2$, $a_{max} + 1$

m/s$^2$} with a step of 0.1 m/s$^2$, $t_B \in \{t_{start}, t_{end}\}$ s with a step of 0.1 s, $j_B \in \{j_{min}$ - 5 m/s$^3$, 0 m/s$^3\}$ with a step of 0.2 m/s$^3$. The $a_{max}$ is the maximum acceleration and $j_{min}$ is the minimum jerk, both within the window of [$t_{start}$, $t_{end}$]. The model is illustrated in Figure 4. The model is designed to return a missing value if it cannot determine all of the aforementioned parameters. The motivation for adapting the grid search parameter ranges to $T_1$, $a_{max}$, and $j_{min}$ is twofold: 1) to ensure that the measured braking response happens after stimulus onset, because we only wish to capture unpremeditated braking in response to the unexpected stimulus (Hankey et al., 2016; Engström et al., 2024); 2) to accommodate the variability in braking kinematics to improve model efficiency. Alternatives of time anchors other than stimulus onset $T_1$ will be addressed in the Discussion.

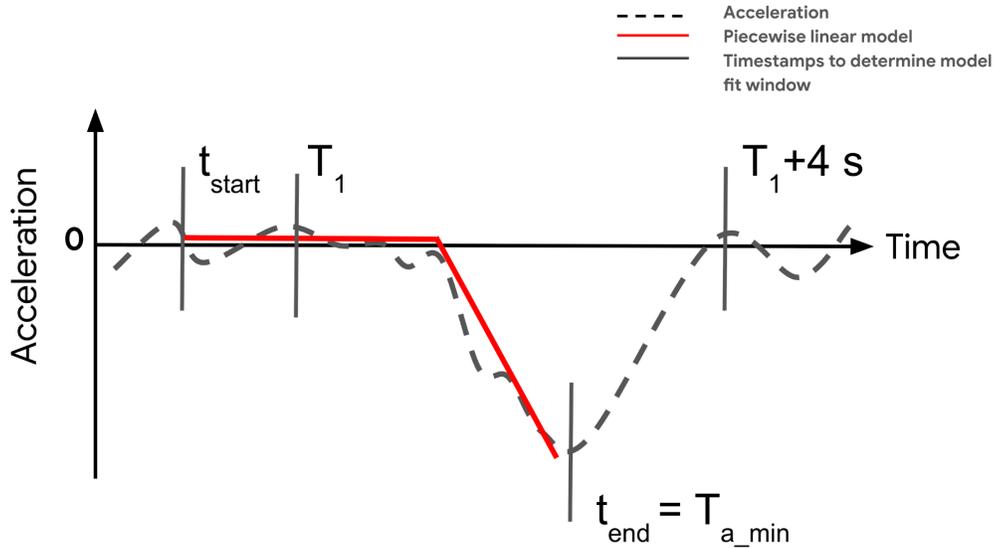

Figure 4. The illustration of the model fit window within which the grid search is defined.

Similar to Markkula et al. (2016), we use the coefficient of determination $R^2$, or the goodness of fit, to select the best fit model (Field, 2009). $R^2$ quantifies the difference between the fitted acceleration by the piecewise linear model and the actual acceleration time series, as the fraction of variance captured by the model, with $R^2$ = 1 representing a perfect fit.

For each event, the model input is the time series of unfiltered acceleration data. The model also uses jerk (i.e., a time derivative of the unfiltered acceleration computed using finite differentiation approximation over two timestamps) to set the range for grid search and manually annotated $T_1$ to define the time window as described above. The output is the $a_0$, $j_B$, $t_B$, and the $R^2$ of this parameter vector of the winning model ranked by $R^2$.

Note that we did not include any filters based on kinematics such as $a_{min}$. This means that the algorithm will try to fit a model as long as it can find an $a_{min}$ within the defined window even when the $a_{min}$ is very small in magnitude such as -1 m/s$^2$ or even positive.

The model was implemented in Google Colaboratory using Python and can be implemented in most other types of programming language.

## Results

The algorithm independently loops through each event that was also manually annotated. The time needed to manually annotate an event is two to five minutes, depending on the complexity of the event. The average time for the algorithm (implemented in Python) to estimate brake onset is less than ten seconds. The differences, or deviations between the estimated and manually annotated brake onset were computed as *model estimation - manual annotation*. A positive difference means that the brake onset estimated by the model is later than the manually annotated response onset and a negative deviation means the algorithm computed an earlier brake onset. This computation would yield missing values in two conditions: 1) when the algorithm cannot determine all of the parameters and 2) when the algorithm outputs a numeric vector but the manual annotation indicates that there was no braking maneuver. Results for these events with missing values are presented separately below. Furthermore, three events were removed due to part of the time series data being incomplete.

### Overall comparison between model estimation and manual annotation

After removing the missing values in the differences and events with incomplete time series, a total number of 190 events were analyzed and distribution of the deviations are shown in Figure 5.

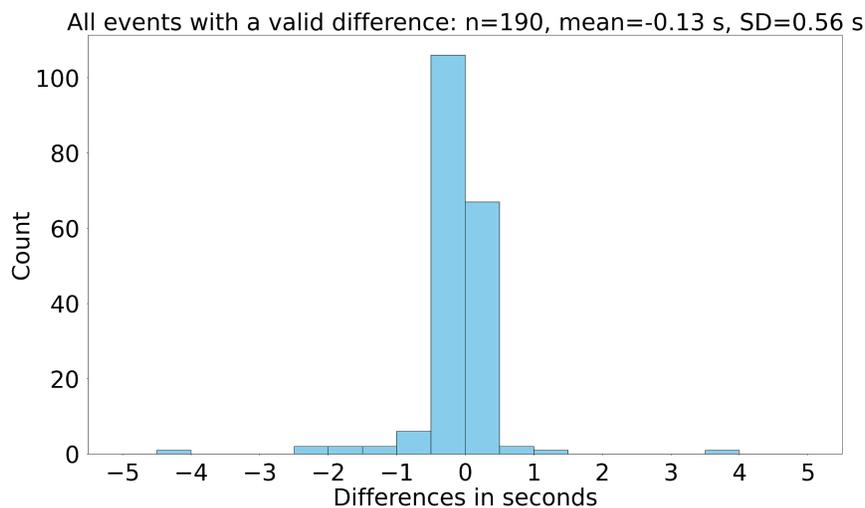

Figure 5. The differences between model estimation and manual annotation of the brake onset. Positive values mean model estimation is later than manual annotation and negative values mean model estimated an earlier brake onset.

The result shows that the deviations between model output and manual annotations are small with 91.1% of the deviations are less than or equal to 0.5 s and 84.2% if using a 0.3 s threshold. The distribution is slightly skewed to the left meaning that the algorithm tends to estimate an earlier brake onset than manual annotation.

We specifically analyzed the 17 events with a difference greater than 0.5 s. The event with the largest deviation of -4.4 s was the one where the responder performed a very late and abrupt response (See Figure 6). The responder was also slowing down for quite some time prior to the actual but late response. Additionally, the acceleration signal was noisy which posed further challenges to the algorithm. The best fit model of this event has a low $R^2$ of 0.56. It is worth noting that this is a challenging event for humans to annotate as well, given the initial slowing down maneuver and noisy signal. Similarly, due the complexity of city driving and changing scenario dynamics, we observe evasive brakings that occurred after an intentional initial slowing down, or complex braking characteristics such as stepwise braking where the evasive braking gradually ramps up to become a harsher maneuver, or multiple consecutive braking maneuvers with a similar peak deceleration magnitude. These complicated, yet natural, characteristics of evasive braking pose difficulties not just to the algorithm, but also to the manual annotation as humans would heavily rely on video data to provide contextual information. These phenomena contribute to 10 of the 17 events with a deviation greater than 0.5 s.

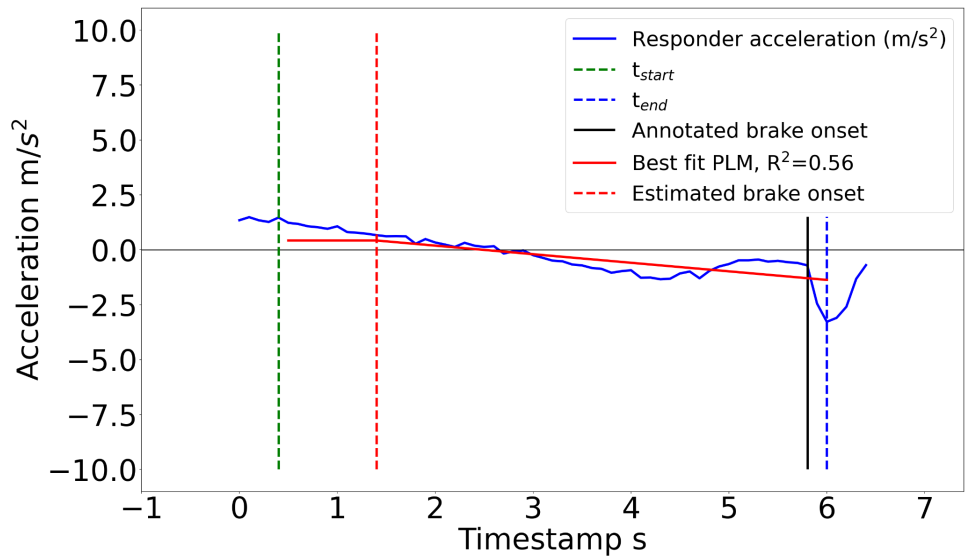

Figure 6. Time series of responder acceleration signal and the best fit PLM of the event with a large deviation of -4.4 s between model estimation and manual annotation.

For the rest of the events, the responder in one event did perform an evasive braking; however, it was outside the model fit window and thus not captured by the algorithm. The responder acceleration signals were noisy in the other five events which negatively affected the model fit.

Model fit in crashes

The dataset contains seven crashes in total and one event was removed due to partially missing time series data. The responders in the remaining six events all braked. Although the impact from a crash could drastically change the trajectory and introduce unrealistic noise to acceleration signals such as a large jerk, we only used the time series until 0.2 s before the crash point and this improved the model fit (see Figure 7).

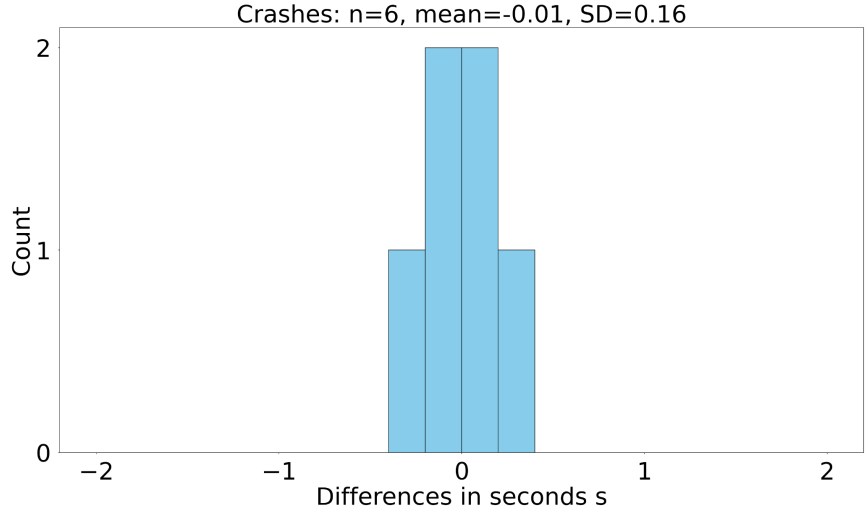

Figure 7. The differences between manual annotated and estimated brake onset in the six crashes.

The impact of a crash on the time series date is demonstrated in Figure 8, where the crash introduced a huge jump in acceleration signal that led to an unrealistic strong acceleration of almost -100 m/s². After removing the time series data after 0.2 s prior to the crash onset, we retained only the acceleration signals that are relevant to responder's evasive braking. The algorithm would have computed an unusually high jerk $j_B$ for the above event if we did not cut off the acceleration signals.

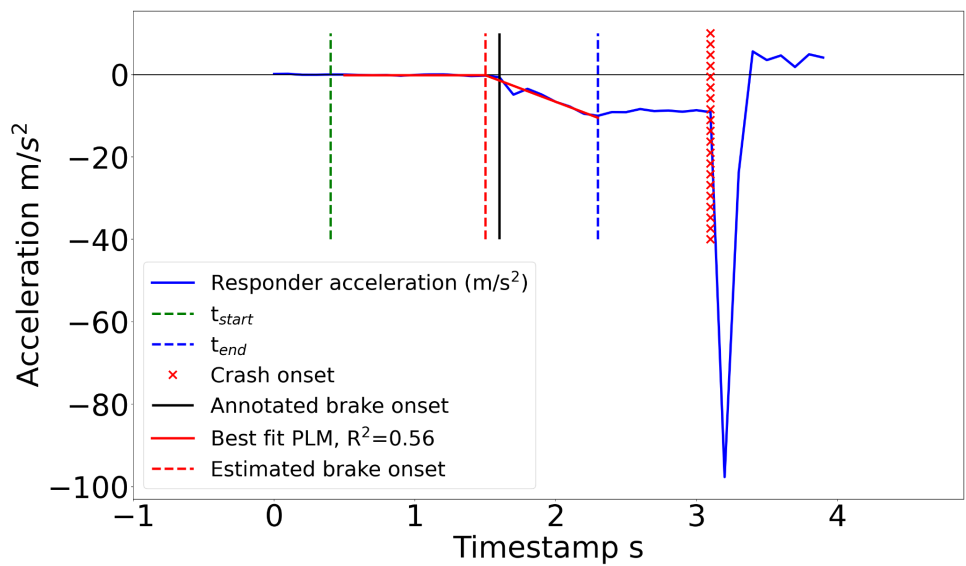

Figure 8. An example of a crash event. The responder acceleration was unrealistically impacted by the crash after the crash onset (red-cross vertical line). The model was able to capture the correct brake onset after removing the acceleration signals shortly before the crash onset.

In general, collision events should be handled with extra caution. Since collisions are rare, manual inspections of these events can benefit an accurate brake onset estimation without introducing too much manual effort.

## Model fit in different road user groups

The responders in the dataset were passenger cars, motorcycles (MC), bicycles (CLC), and pedestrians (Ped), or vulnerable road users (VRU), and they are less common in the dataset (see Figure 9). Due to the small sample size of these responders, we aggregated them into a non-car group. Similar to the reasons of outliers presented earlier, the outliers with greater than 0.5 s deviation are because of noisy acceleration signals or agents performing stepwise braking or multiple slowing down maneuvers in addition to the actual evasive braking. The most common category was passenger cars (See Figure 10).

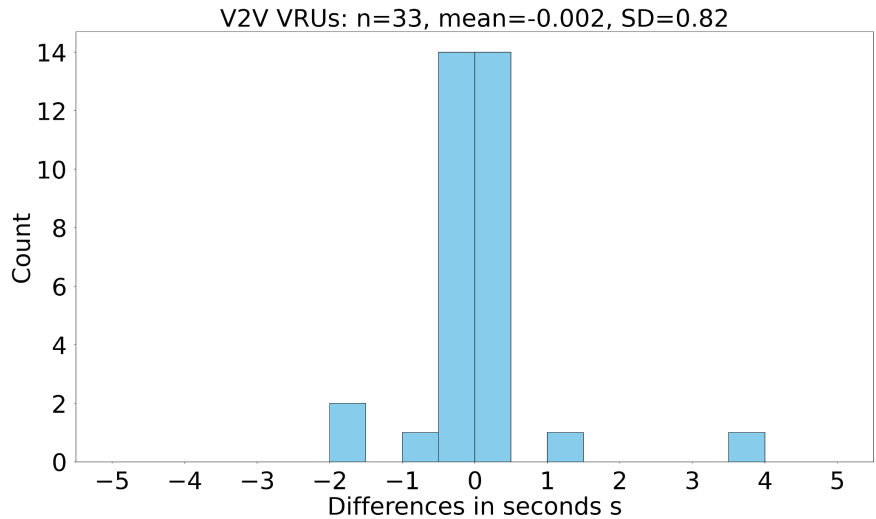

Figure 9. The differences between model estimation and manual annotation among non-car road user groups.

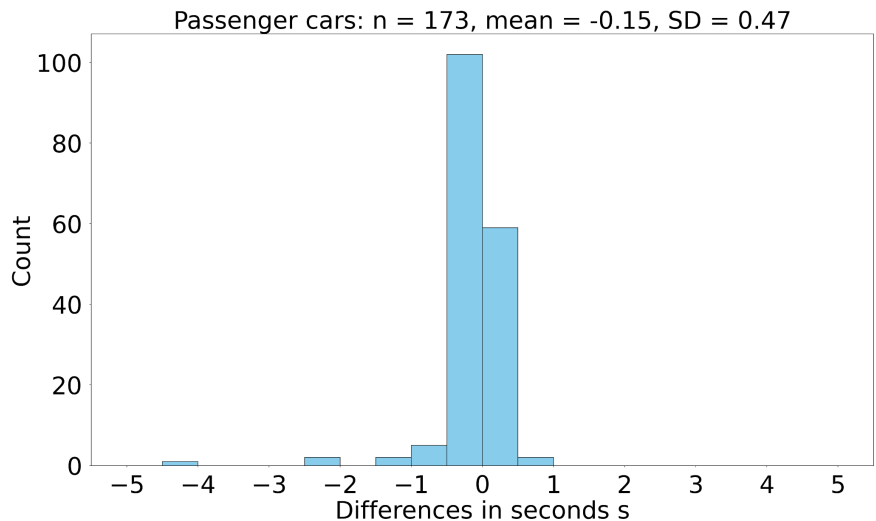

Figure 10. The differences between model estimation and manual annotation among passenger cars.

The differences in non-passenger car and passenger car groups have similar distributions which suggest that the algorithm is applicable to different road user types.

Model fit in different scenario types

This study analyzed data in eight different scenarios (Table 2). To examine if the model performs differently in different scenarios, we further collapse the scenarios into three types: 1) V2V F2R, 2) V2V Non-F2R (i.e., intersection, lateral, opposite direction, secondary crash), and 3) VRU (Vulnerable Road Users, i.e., motorcycle, cyclist, and pedestrian). The distribution of differences among the further collapsed scenario types are shown in Figure 11. Since the VRU scenario corresponds to the VRU road user type in Figure 9, this scenario is not included in Figure 11.

The plot shows similar distributions between V2V F2R and Non-F2R scenarios: majority estimations are less than 0.5 s different from manual annotations that take the contextual information into account. In scenarios involving a VRU (see Figure 9), the estimations have a slightly larger proportion of events with greater than 0.5 s difference when compared to manual annotations.

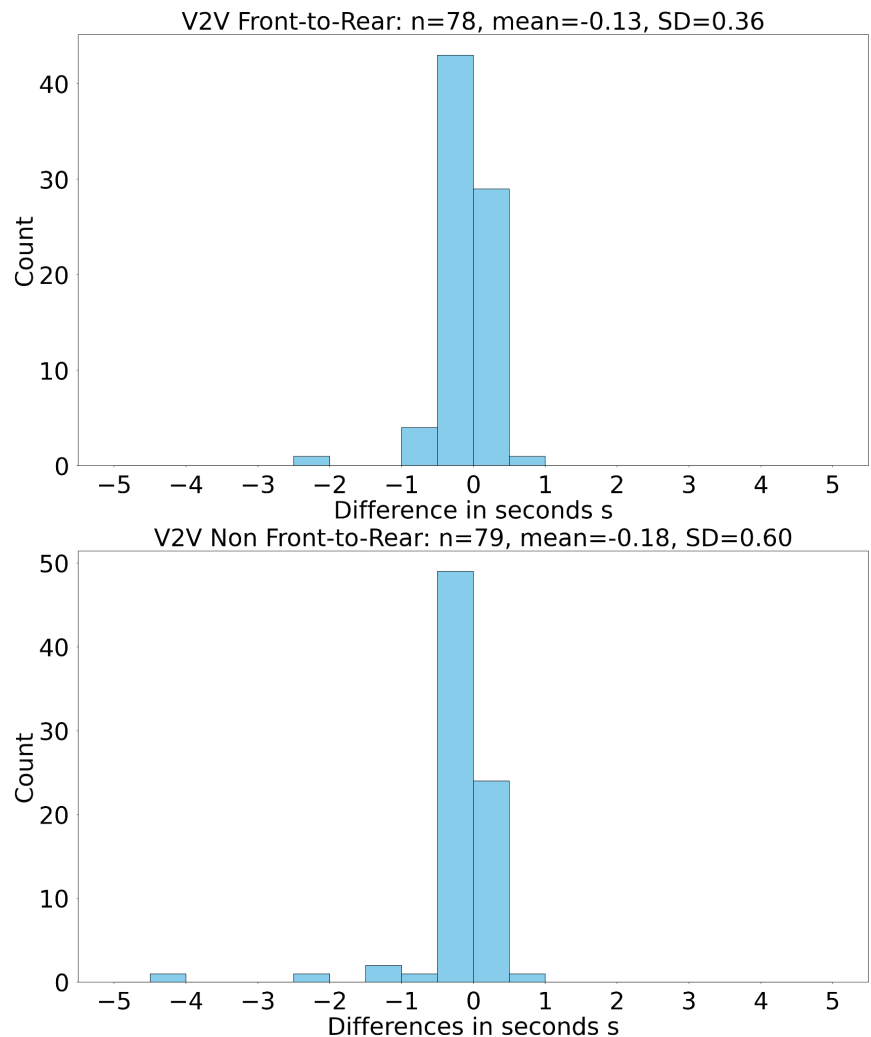

Figure 11. The differences between model estimation and manual annotation in different scenario types.

Confidence metrics for results selection

The main goal of the proposed model is to automatically estimate brake onset in traffic conflicts including near-crashes and crashes. Admittedly, the model can perform well in many events but not all events such as when the road user performs an irrelevant braking maneuver within the model fit window. Therefore, in order to apply the model on large scale datasets and select accurate estimation, we need a confidence metric to identify events where the model performs well. Depending on the accuracy and efficiency requirements of the use case, users can choose to only keep the events with good model performance or to have the events where model likely failed manually annotated. The model output was the best model among the grid search and was ranked by $R^2$ or the goodness of fitness. The larger the $R^2$ is, the better the fit to the acceleration data. The model outputs the parameters that yield the highest $R^2$ possible, thus it is natural to use these highest possible $R^2$ as the confidence metric to determine the extent to which the model yields reliable results (Markkula et al., 2016).

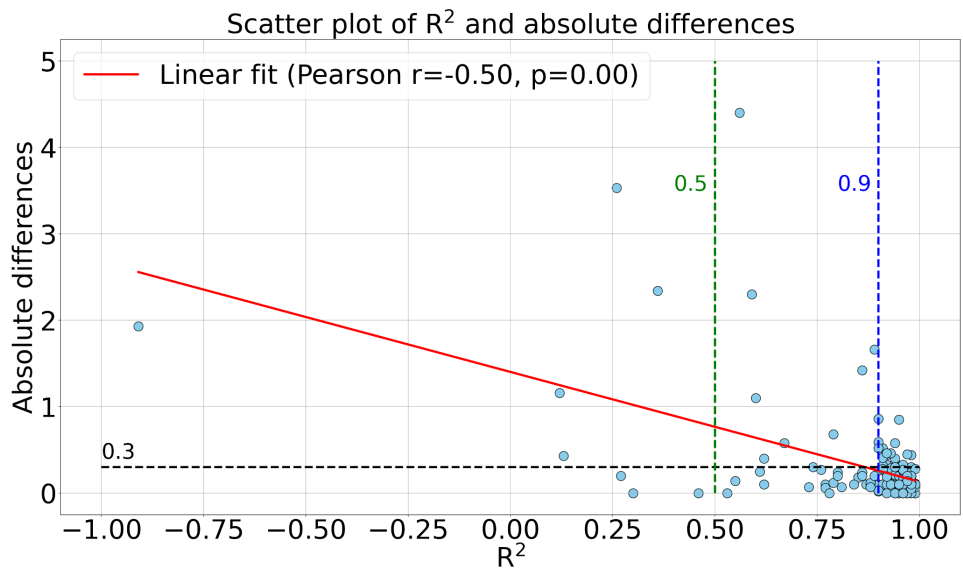

Figure 12. Scatter plot of $R^2$ and the absolute difference between manual annotations and model estimations and the red line is a linear regression fit between them.

The linear regression model shows that there is a statistically significant (p-value=0.00) relationship between $R^2$ and the absolute differences between manual annotation and model estimation and the correlation is negative (Pearson's r=-0.5): the absolute differences decrease as $R^2$ increases. Since the goal is to select "good" estimations instead of predicting the actual difference, a regression model is not suitable. For example, we can define an estimation as "good" if the absolute difference is smaller than a threshold (e.g., the events below the black dashed line in Figure 12) and define a good model performance if the $R^2$ is greater than a threshold (e.g., the events to the right of the green or blue dashed line in Figure 12). Given a constant threshold of the difference criterion, we can select results by varying the threshold on $R^2$. If we move the $R^2$ threshold to 0.5 (green dashed line in Figure 12), we would cover most of the good estimations; however, we would also select many events with greater than threshold differences. If we move the $R^2$ threshold to 0.9 (blue dashed line in Figure 12), although we avoid almost all events with large deviations, it also excludes some good estimations. Balancing the inclusion of good and bad estimations is the key here. Thus, we created a classifier model to help users determine a suitable threshold for $R^2$ and use it as the confidence metric.

To select results with good performance, we developed a classifier model. We began by defining *positive events* as those with a high $R^2$ value, indicating a reliable model estimate. Conversely, *potentially negative events* are those with a low $R^2$ value, reflecting lower confidence in the estimated brake onset. We then classified events based on the absolute difference between the estimated and actual brake onset times. Events with absolute differences less than or equal to 0.3 seconds are considered *small-difference events* (i.e., *actually positive events*), while those with differences greater than 0.3 seconds are labeled *large-difference events* (i.e., *actually negative events*). The aim of the confidence metric is to distinguish between actually positive and actually negative events. There is no established standard or prior research to guide the selection of the 0.3-second threshold in this context; it was chosen as a practical error tolerance margin. This threshold can be adjusted depending on the specific application and required accuracy. However, based on empirical observations, we consider 0.5 seconds to be a reasonable upper bound for this threshold. Finally, by setting a threshold for $R^2$, we categorized events as *predicted positive* ($R^2 \geq$ threshold) or *predicted negative* ($R^2 <$ threshold). The classifier model is illustrated in Figure 13.

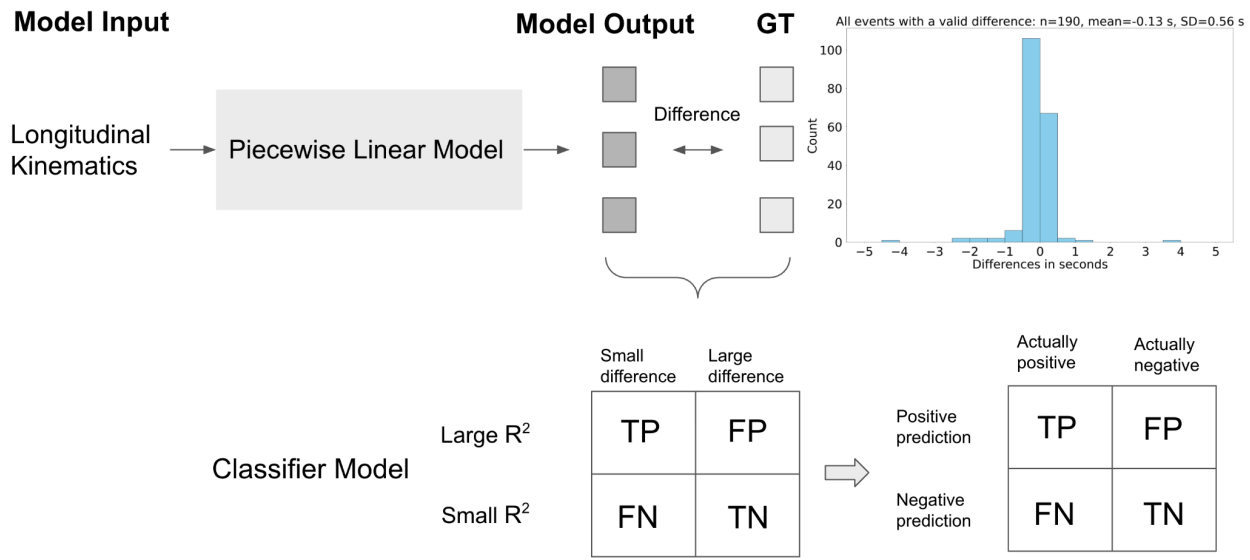

Figure 13. Illustration of piecewise linear model and the construct of confidence metric model.

We want to find a proper threshold for $R^2$ so that it can optimize the coverage of true positives (TP, i.e., events with small differences and large $R^2$) and minimize false negatives (FN, i.e., events with small differences but small $R^2$). For actual negatives, the confidence metric can minimize the capture of false positives (FP, i.e., events with large differences but large $R^2$) and optimize the identification of true negatives (TN, i.e., events with large differences and small $R^2$). We want to control the amount of false predictions while optimizing the amount of true predictions when using a confidence metric to select results. The following metrics were computed to help select a reasonable threshold:

- $Recall\ or\ True\ Positive\ Rate\ (TPR)\ =\ TP\ /\ (TP\ +\ FN)$

- $False\ Positive\ Rate\ (FPR)\ =\ FP\ /\ (FP\ +\ TN)$

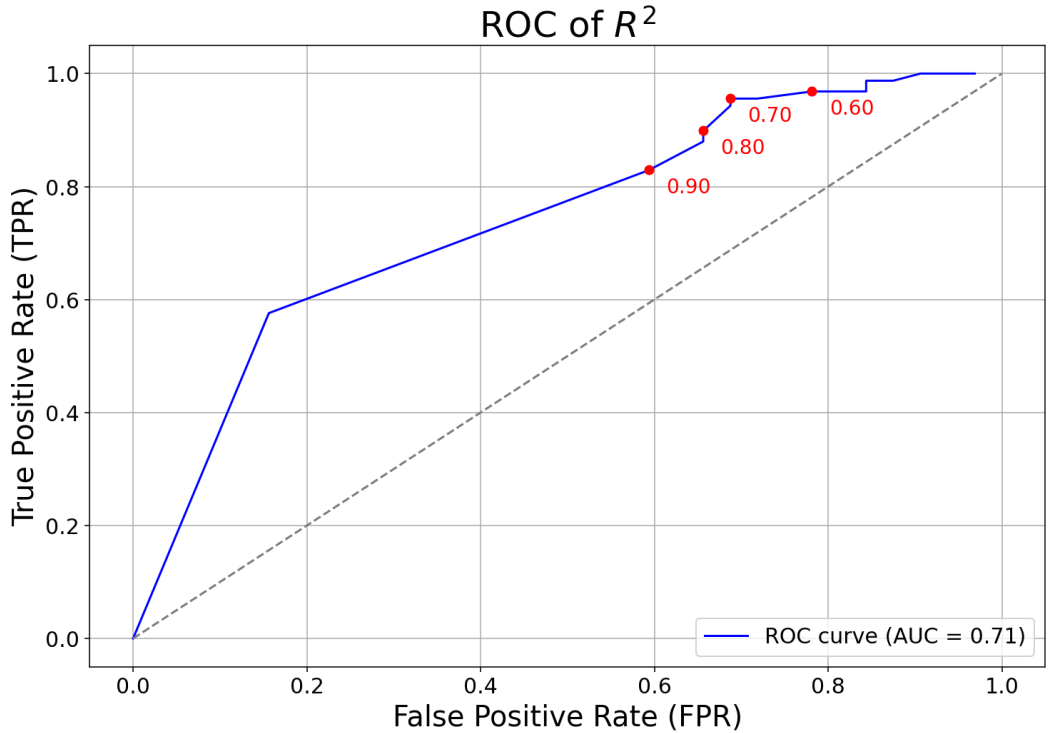

Figure 14. The Receiver Operating Characteristic (ROC) curve (the blue line) and Area Under Curve of $R^2$. The red dots are the few $R^2$ thresholds greater than 0.5: the red numbers are the threshold values and the corresponding FPR is on the X axis and TPR is on the Y axis.

We plotted the receiver operating characteristic (ROC) across $R^2$ thresholds from 0 to 1 with a step of 0.1 for $R^2$ to show the confidence metric performance at different thresholds. We also calculated the Area Under Curve (AUC) to obtain a measure of the general performance of $R^2$ in predicting poor response onset estimations by the model. AUC represents the probability of the confidence metric to correctly distinguish positives from negatives and an AUC of 0.5 (the grey dotted line in Figure 14) represents 50% probability which is the same as a random guess. Figure 14 shows that the confidence metric can distinguish positives (small diff events) and negatives (large diff events) properly as it has an AUC of 0.71. For a specific confidence metric, ROC does not recommend a threshold, instead, it shows the potential trade-offs between TPR and FPR of a given threshold. The intention of analyzing the two potential confidence metrics is to provide a way to automatically select results for rapid estimation of brake onset on large scale data rather than prescribing a threshold to use. A low threshold for $R^2$ would help identify more true positives at the sacrifice of a higher false positive rate while a high threshold would reduce false positive rate but at the sacrifice of a lower true positive rate (e.g., the red dot of 0.6 vs. red dot of 0.9 in Figure 14). The exact threshold selection should be based on whether the users want to prioritize true positive coverage or a low false positive rate. For example, if the goal is to have an accurate evaluation, the threshold should be on the higher end and pass the negative events to manual evaluation.

Below is the distribution of the $R^2$ of the best fit model of each event (Figure 15). Examples of when the algorithm made accurate estimations or failed to capture the correct evasive brakings were described in the figure caption.

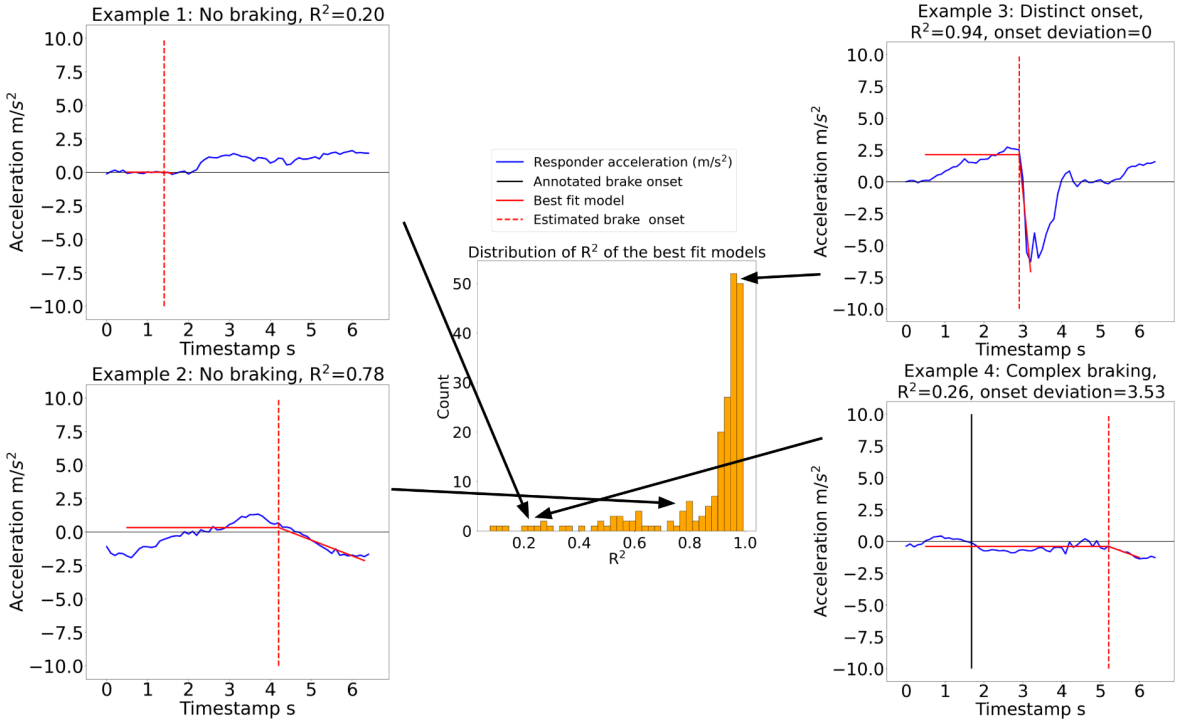

Figure 15. Distribution of the positive $R^2$ of best fit models. Example 1 is an example of when a responder did not brake and the algorithm output a valid parameter vector with a low $R^2$ due to noise and fluctuations in acceleration signal. The $a_{min}$ is only -0.15 m/s$^2$. Example 2 shows when a responder did not brake and the algorithm captured a slowing maneuver within the model fit window. The video data confirmed that the responder only braked after already longitudinally passed the initiator and the behavior was not in response to the conflict. Example 3 is when a responder performed a single brake ramp braking that had a clear onset. The best fit model has a high $R^2$ of 0.94. The onset deviation between model estimation and manual annotation is 0. Example 4 shows a large onset deviation event due to the complexity in responder's braking maneuver. Based on context provided by video, the responder initiated the braking quite early (indicated by the vertical black line); however the braking was light and had some fluctuations. The braking only became harsher a few seconds after the manually annotated brake onset which was captured by the algorithm. This was a low speed event and the initial traveling speed of the responder was about 7 m/s or 15.7 miles/hour.

Events with missing deviation values

The comparisons between model estimated and manually annotated brake onset so far was based on the deviations between the two numeric values; however, as discussed earlier the deviation values could be missing for three reasons: 1) the algorithm could not find a valid value for any of the parameters; 2) the responder did not brake but steered (i.e., no braking); 3) the responder did not respond at all (i.e., no response). The current dataset contains 26 such events and is summarized in Table 3 and Table 4.

In the motorcycle event where the responder did not respond at all and the algorithm detected a braking maneuver (See Table 3), the initiator braked for a red light and traffic ahead while the motorcyclist was following very closely and lane splitting which led to a conflict. The initiator slightly steered which gave the motorcyclist more space to pass without any response. The motorcyclist braked for the red light shortly after longitudinally passing the initiator which was captured by the algorithm and mistaken as its braking response. In other *no response* events, the algorithm incorrectly detecting a brake onset was due to 1) irrelevant slowing down as the responder was no longer interacting with the initiator; or 2) fluctuations in acceleration signals. None of the *no response* events have a $R^2$ that is greater than 0.6. The same two reasons contribute to the fourteen events where the algorithm detected a braking maneuver while the responder only responded by swerving.

Table 3. Reason for missing deviation values by agent type. *Note that the one event where the model output was missing was also an event where the responder did not brake.*

|  | Passenger cars | Motorcycles | All agent types |
|---|---|---|---|
| **No model output** | 1 | 0 | 1 |
| **No braking** | 14 | 0 | 14 |
| **No response** | 11 | 1 | 12 |
| **Total** | 25 | 1 | 26 |

When looking at scenario types (Table 4), most of the no response and no braking happened in the V2V lateral scenario (see Figure 1 for conflict typology). More than half of the responders did not brake but performed a lateral maneuver to avoid the collision.

Table 4. Reason for missing deviations by scenario type. *Note that the one event where the model output was missing was also an event where the responder did not brake.*

|  | V2V VRU | V2V F2R | V2V Opposite Direction | V2V Lateral | All scenario types |
|---|---|---|---|---|---|
| **No model output** | 1 | 0 | 0 | 0 | 1 |
| **No braking** | 1 | 1 | 1 | 11 | 14 |
| **No response** | 1 | 3 | 0 | 8 | 12 |
| **Total** | 2 | 4 | 1 | 19 | 26 |

One noticeable failure of the algorithm was that it sometimes mistook noise and fluctuations in the acceleration signal as a braking maneuver (See Figure 14 example 4) when there were no braking maneuvers at all. Urban and city driving is dynamic and drivers often do not drive at a constant speed due to various reasons. Although we did not filter out events based on magnitude of small minimum acceleration, it could help identify no braking and no response events from the results. Using a -0.3 m/s$^2$ threshold and filter out events where $a_{min}$ >= threshold, we are able to identify nine no braking or no response events. However, when we use a -0.4 m/s$^2$ threshold we captured thirteen no braking or no response events but we would also mistakenly filter out two valid events where the responder did brake and the braking magnitude was very small because of low speed.

## Discussions and Conclusions

To efficiently estimate brake onset in traffic conflicts for human drivers and ADS, we developed 1) a consistent manual annotation approach; 2) a simple and easy to implement algorithm that automatically estimates brake onset at scale; 3) a confidence metric that was validated on real conflicts to guide users to select results automatically. The method enables efficient analyses and evaluations of evasive braking maneuvers in naturalistic driving logs and simulated ADS driving data, as well as other types of data that contain time series of longitudinal kinematics. A previous study developed a three-piece piecewise linear model and analyzed braking responses in rear-end conflicts (Markkula, et al., 2016). In the current study, we modified this three-piece model, which was designed to model the full braking profile, to a two-piece model optimized for identifying the brake onset time stamp, and customized the parameter grid search based on the kinematics of individual events. These modifications thus focus on optimizing brake onset estimation and efficiency, for example, having less parameters and customizing grid search contribute to better brake onset estimation (e.g., $R^2$ would not be affected by fluctuations in final steady state acceleration ) and efficiency (e.g., a smaller search grid). While the previous research studied rear-end conflicts, we applied the model to eight conflict scenarios including low speed events (see Table 2). Our results showed that this larger variety of conflict types did not negatively impact the model performance (Figure 11). Markkula et. al (2016) included responses of trucks and buses in the analysis which the current study does not. This is likely due to the current events being sampled from dense urban areas. We believe the model would be able to properly estimate brake onsets of trucks or buses because they have the same braking mechanism as passenger cars. Our study expanded the agent types to cover vulnerable road users including motorcycles, bicycles, pedestrians on foot, and pedestrians using micromobility devices. In summary, our model proves to be able to estimate brake onsets for different types of road users in a wide range of scenarios.

There are three other advantages of the proposed model when using it to automatically estimate brake onset for a variety of analyses such as evaluating response timing of ADS and developing behavior benchmarking models (Engström et al., 2024). First, this is a minimalist model that in principle only requires a single input signal: longitudinal acceleration and estimates only four parameters. In the current analysis, jerk, derived from acceleration time series, was used as an additional input to determine parameter range. Second, all four parameters are highly configurable. In our study, we proposed a customized grid search based on kinematics of individual events which proved to be efficient and reliable. The search grid for parameters and granularity of it (e.g., step size) can be configured in a way that is suitable for the specific data, use case, computational power, and desired accuracy. Finally, we proposed to use $R^2$ as a confidence metric and evaluated the model performance against ground truth annotations. The analysis shows that the higher the value of the metric, the more accurate the response onset estimation is. The $R^2$ is able to classify the results into small differences and large differences compared to manual annotation with a reasonable performance (AUC = 0.71). The metric enables automatic selection of results based on requirement on accuracy and thus eliminates the need for manual annotation of large scale dataset when fast iteration is desired. If a higher accuracy is desired, the manual guidelines can be used to evaluate the events that are flagged by a lower than threshold $R^2$.

There are some known limitations in the model and the overall approach to estimate response onset. Most notable is the exclusion of swerving maneuvers. Despite being generally less common compared to braking responses on city streets, steering maneuvers are important to avoid collisions especially at higher speeds such as in highway scenarios. We have attempted to apply the model on time series of lateral acceleration and yaw rate to measure the swerving onset; however, the attempt was not successful due to several reasons: 1) lateral kinematic signals are noisier than longitudinal signals and are typically sinusoidal (i.e., wave-shaped) which is not well captured by the current model; 2) the shape of swerving maneuvers varies a lot depending on the travelling speed. Prior research shows that some simpler metrics such as maximum lateral acceleration might be useful for quantifying and evaluating swerving maneuvers (Terranova et al., 2024); however, more research is needed to develop a comprehensive set of automatic measures to detect evasive swerving.

The second limitation is that the model does not account for contextual information as it relies solely on kinematics. As discussed in Engström et. al (2024), and exemplified in Figure 6, drivers might have been slowing down prior to the actual evasive braking maneuvers. And in those situations, the model might capture the onset of the initial planned braking maneuver instead of the onset of actual evasive braking.

The third limitation is the single brake ramp assumption of the piecewise linear model. This limitation mostly affects vulnerable road users who engage their feet on the ground to brake or stop such as runners and skateboarders. The discrete contacts with the ground introduce noise in the signal as well multiple brake ramps (Terranova et al., 2024). Although we did not observe the model performing worse due to these reasons in the current analysis (likely due to the limited sample size of VRU events in our dataset), future applications of this model should take extra caution when analyzing road users who use their feet to brake or stop. Related to the fluctuations introduced by foot touching ground, noise in the signal can also negatively affect the results and users should properly smooth the data.

Although not a limitation, it is worth noting that the practical application of the model presupposes the users have a way to identify conflicts in log data. This is a complicated problem and is not something we addressed in the paper. Surrogate measures can be used to discover near-crashes from raw data such as measures of hard braking and required deceleration (Perez et al., 2017) and the can further enhanced by computational, cognitive science-based, metrics such as surprise (Dinparastdjadid et al., 2023). It is also worth noting that this paper used manually annotated $T_1$ (Engström et. al, 2024) to determine the model fit window. The model fit window can be determined using any anchor time as long as it captures a time period during which evasive maneuvers are most likely to happen. For example, surrogate measures such as minimum post-enchroachment time or minimum time to collision can be used to encapsulate a potential conflict window for model fit.

In summary, we developed an efficient and versatile evaluation process for evasive braking in traffic conflicts that consists of an automated algorithm, confidence metric, and a manual evaluation guideline. We believe this method provides a unified solution to estimate brake onset using kinematic signals that can be used for many applications in the context of safety evaluation for Level 4 ADS, and road safety analysis in general. With this method, for example, not only can we benchmark braking response timing of Level 4 ADS (either the log or simulated responses) on human data, but it also enables comparisons across data sources, such as comparing simulated Level 4 ADS performance in long tail events to human driving performance in driving simulator study data.

To achieve a comprehensive evaluation process for evasive maneuvers, key areas for further research include analysis and modeling methods that account for steering behaviors and complex braking maneuvers, e.g., braking with multiple brake ramps.

Terranova, P., Liu, S. Y., Jain, S., Engström, J., & Perez, M. A. (2024). Kinematic characterization of micro-mobility vehicles during evasive maneuvers. Journal of Safety Research, 91, 342-353.

Venkatraman, V., Lee, J. D., & Schwarz, C. W. (2016). Steer or brake?: Modeling drivers' collision-avoidance behavior by using perceptual cues. *Transportation research record*, 2602(1), 97-103.

Victor, T., Dozza, M., Bärgman, J., Boda, C. N., Engström, J., Flannagan, C., ... & Markkula, G. (2015). Analysis of naturalistic driving study data: Safer glances, driver inattention, and crash risk (No. SHRP 2 Report S2-S08A-RW-1).

Waymo (2025, July). *The Waymo Driver has officially driven 100 million fully autonomous miles on public roads* [Post]. LinkedIn.
https://www.linkedin.com/posts/waymo_100m-autonomous-miles-activity-7350871765817970689-LGvT/23